\documentstyle[12pt]{article}
\newcommand{\be}{\begin{equation}}
\newcommand{\ee}{\end{equation}}
\newcommand{\ba}{\begin{eqnarray}}
\newcommand{\ea}{\end{eqnarray}}
\newcommand{\ft}{\footnote}
\newcommand{\nl}{\newline}
\newcommand{\al}{\alpha}
\newcommand{\bt}{\beta}
\newcommand{\ga}{\gamma}
\newcommand{\pr}{\prime}

\begin{document}

\begin{flushright}
QMW-PH-97-22
\end{flushright}

\begin{center}

\Large{\bf On Mirror Symmetry for Manifolds of Exceptional Holonomy.}

B.S.Acharya\ft{r.acharya@qmw.ac.uk \nl Work Supported by PPARC.},

{\it Department of Physics, Queen Mary and Westfield College, Mile End Road,
London. E1 4NS. U.K.}

\end{center}

\begin{abstract}

We consider Type II string theories on ${\bf T^n}/{{\bf Z_2}^m}$ Joyce
orbifolds. This class contains orbifolds which can be desingularised to
give manifolds of $G_2$ $({\bf n}$$=$$7)$ and $Spin(7)$ holonomy
$({\bf n}$$=$$8)$. In the $G_2$ holonomy case we present two types
of $T$-duality transformation which are clearly generalisations of
the $T$-duality/mirror transformation in Calabi-Yau spaces. The first
maps Type IIA theory on one such space from this class to Type IIB
theory on another such space. The second maps Type IIA (IIB) to Type IIA
(IIB). In the case of $Spin(7)$ holonomy we present a
$T$-duality transformation which maps Type IIA (IIB) theory on one
such space to Type IIA (IIB) on another such space. As orbifold
conformal field theories these $T$-dual target spaces are related via
the inclusion/exclusion of discrete torsion and the $T$-duality is
proven to genus $g$ in string perturbation theory. We then apply a
Strominger, Yau, Zaslow type argument which suggests that manifolds
of $G_2$ holonomy which have a ``mirror'' of the first (second)
type admit supersymmetric ${\bf T^3}$ (${\bf T^4}$) fibrations and that
manifolds of $Spin(7)$ holonomy for which a mirror exists admit fibration
by supersymmetric $4$-tori. Further evidence for this suggestion is
given by examining the moduli space structure of wrapped D-branes.
\end{abstract}
\newpage
\section{Introduction.}

Mirror symmetry has proven to be a powerful and fascinating area
of string theory (for a review see \cite{Mirror}). 
Mirror symmetry is at present only defined for
string theory on Calabi-Yau spaces. Recent work \cite{sv,pt,ba1}
has suggested the possible existence of an analagous symmetry for
string theory on manifolds of exceptional holonomy.
This evidence is perhaps not surprising since
manifolds of exceptional holonomy share many properties with Calabi-Yau
manifolds. They admit Ricci flat metrics and covariantly constant forms
leading to the important fact that they give supersymmetric compactifications
of string theory and $M$ theory. Another remarkable general feature is that 
Euclidean D-branes wrapped around supersymmetric
cycles in these manifolds have world-volume theories which are
cohomological \cite{dtop,wv,bks,ba3,ba4}. 
This latter feature is almost certainly related
to what we will have to say in this paper, but will not be discussed here.

String theories on orbifold limits of such manifolds can provide
a simple framework in which to study a variety of physical and mathematical
phenomena which often persist as one moves away from the orbifold limit.
This is due to the fact that strings remarkably ``know how to'' blow up
the orbifold. We will exploit these simplifications here.

The purpose of this paper is to provide evidence for the
existence of a symmetry in Type IIA/IIB string theory compactified
on manifolds of exceptional holonomy. As we will see, this symmetry
is very much analogous to the mirror symmetry of Type II strings on
Calabi-Yau $n$-folds. 

In the case of Calabi-Yau threefolds, Strominger, Yau and Zaslow 
(SYZ)
considered some non-perturbative implications 
of mirror symmetry \cite{syz}.\newline
By considering the mirror BPS soliton spectra in the two theories
they argued that mirror symmetry is $T$-duality. The analogue of their
argument for Calabi-Yau $n$-folds was given in \cite{mor}. The outcome of
the SYZ argument is that Calabi-Yau $n$-folds which have a mirror admit
supersymmetric ${\bf T^n}$ fibrations. 
Mirror symmetry is then $T$-duality on these
supersymmetric fibers.

Owing to the simplifications in studying orbifolds, one well understood example
of mirror symmetry is the pair of ${\bf T^6}/{{\bf Z_2}{\times}{\bf Z_2}}$
orbifolds first constructed in \cite{disc}. This pair of orbifolds differ
by the inclusion/exclusion of discrete torsion\ft{Discrete torsion was defined
in \cite{vafa}.}. In this case, for each
choice of orbifold isometry group, there is only one consistent choice for the
discrete torsion leading to the fact that there are only two such orbifolds,
one without discrete torsion and the other with. As pointed out in \cite{disc}
this pair of orbifolds have mirror Hodge numbers. Let us
denote these two target spaces as ${\bf X}$ and ${\bf Y}$.
This pair of orbifolds
were further studied in \cite{vw} where it was shown that Type IIA theory
on ${\bf X}$ is equivalent to Type IIB theory on ${\bf Y}$. This
equivalence was proven to all orders in string perturbation theory.
Moreover the
mirror symmetry transformation between these two theories was shown to
be $T$-duality. This example thus provides strong evidence that the
SYZ argument is correct\ft{Further aspects of this mirror pair were studied
in \cite{oz} in light of \cite{syz}.}. 

In \cite{ba2} we tested the SYZ argument for Calabi-Yau fourfolds
by studying ${\bf T^8}/{{\bf Z_2}^3}$ orbifolds. The mirror
transformation is also $R \rightarrow 1/R$ $T$-duality, but this time on
supersymmetric ${\bf T^4}$ fibers. In this case also, it is the turning
on of discrete torsion via the $T$-duality transformation which
``produces'' the mirror target space. As in \cite{vw} equivalence between
these mirror theories can be checked to genus $g$ in string perturbation
theory. 

In \cite{J1,J2,J3} Joyce showed that manifolds with $G_2$ and $Spin(7)$
holonomy can also be constructed by desingularising ${\bf T^n}/{{\bf Z_2}^m}$
orbifolds, where $n$ $=$ $7$ and $8$ for $G_2$ and $Spin(7)$ holonomy
respectively. We studied Type II theories on such orbifolds (for $n$ $=$ $7$)
in \cite{ba1} where we proved that Type IIA theory on a given orbifold
from this class is equivalent to Type IIB theory on another orbifold
from the same class. This verified conjectures made in \cite{sv,pt}.
Again, these examples follow the same pattern: the transformation between
the two theories is $T$-duality on supersymmetric tori, they differ
via the inclusion/exclusion of discrete torsion and the equivalence
can be checked to all orders in string perturbation theory. In this sense,
this transformation between Type IIA and IIB theories on spaces of $G_2$
holonomy can be regarded as a direct generalisation of mirror symmetry for
Calabi-Yau $n$-folds.
In this paper we will further
develop the notion of a ``mirror'' symmetry for
manifolds of exceptional holonomy. The plan of the paper is as follows.

Motivated by the results which follow we first give a simple definition
of mirror manifolds in Type II string theory. This definition is a very
general one and, for the cases which we study here, is likely to be
replaced with a much more specific conjecture in the future. The results
of this paper are independent of this definition, but we make this
definition for two reasons: $(i)$: it allows a simplified presentation
of the
results; $(ii)$: since, in the examples studied here, a
symmetry analogous to mirror symmetry for Calabi-Yau
spaces exists, it deserves
to have a name. For obvious reasons we choose to call it
exceptional mirror symmetry, or just mirror symmetry when the context is
obvious. Throughout this paper a generic pair of
mirror target spaces will be denoted
by ${\bf X}$ and ${\bf Y}$ repsectively.

In section three we study Type II strings on ${\bf T^7}/{{\bf Z_2}^n}$
orbifolds. This class of orbifolds includes those which can be blown up
to give manifolds of $G_2$ holonomy \cite{J1,J2}. We describe two types of
$T$-duality/mirror transformations which map between pairs of target
spaces in this class. The first are of the type studied in
\cite{ba1} and map Type IIA on one target space ${\bf X}$
from this category to
Type IIB on the mirror ${\bf Y}$, 
where the mirror is defined in the previous section.
The second type of transformation maps Type IIA on ${\bf X}$
to Type IIA on ${\bf Y}$ and similarly maps IIB to IIB.

In section four we study Type II theories on ${\bf T^8}/{{\bf Z_2}^n}$
orbifolds, some of which admit a resolution to manifolds of $Spin(7)$
holonomy. We present a $T$-duality/mirror transformation which maps
Type IIA (IIB) theory on one such space (${\bf X}$) to Type IIA (IIB) theory
on the mirror ${\bf Y}$.

In section five we consider the Strominger, Yau, Zaslow argument \cite{syz}
applied to all these three cases. Even though the results are not as
strong as in the Calabi-Yau cases, the results suggest that manifolds of $G_2$
holonomy with a mirror of the first type admit supersymmetric ${\bf T^3}$
fibrations and that those with a mirror of the second type admit
supersymmetric ${\bf T^4}$ fibrations. Similarly, it is conjectured that
manifolds of $Spin(7)$ holonomy for which a mirror exists admit supersymmetric
${\bf T^4}$ fibrations.

In section six we give further evidence for the conjectures of section five
by considering the geometry of the classical orbifolds themselves. We first
point out that all the orbifolds we have discussed indeed have the predicted
fibration structure. We then show that the moduli space of the relevant 
D-brane
wrapped around the corresponding $p$-torus is also an orbifold in the class
we are considering.
\section{A Definition of Mirror Symmetry.}

As we have discussed in the introduction, we will describe
in this paper a symmetry analogous to mirror symmetry
for Calabi-Yau $n$-folds in Type II string theories on manifolds of
exceptional holonomy. In order to make the presentation of these results
clearer as well as to give this symmetry a name it is useful to give
a definition of such a symmetry here.
Mirror symmetry in the conventional 
sense applies to two Type II theories compactified on
different manifolds of $SU(n)$ holonomy. 
Both compactifications yield isomorphic physics.
The perturbation theories on the mirror manifolds
are also directly related. Motivated by these observations,
we will study the following
circumstances: Type IIR 
string theory (where R can be A or B) compactified on a
manifold ${\bf X_d} $ yields completely isomorphic physics to Type IIS
string theory (S can be A or B) 
compactified on the manifold ${\bf Y_d}$. The two perturbation theories
are directly related under the map from one theory to the other. Here, R and S
can be either A or B depending upon the circumstances, and
in general ${\bf X}$ is a topologically different manifold from ${\bf Y}$.
Both manifolds
are $d$-dimensional.

{\it Definition:} When two theories satisfying the above criteria exist
we will say that the two theories are mirror, and that ${\bf X}$ and
${\bf Y}$ are mirror manifolds.

We will now restrict our 
attention to the
case when both mirror \nl 
compactifications preserve some fraction of
the supersymmetries of the ten-dimensional theory. This means that 
both ${\bf X_d}$ and ${\bf Y_d}$
are Ricci flat manifolds of ``special'' holonomy
\ft{By ``special'' we mean $SU(n)$, $G_2$ or $Spin(7)$.}.
Both ${\bf X}$ and ${\bf Y}$
must have the same holonomy as we do not expect two theories 
with different numbers of supersymmetries to yield isomorphic physics.
We will be interested in the cases when both manifolds have holonomy $G_2$
or $Spin(7)$. In these cases, if the need arises to differentiate between
the mirror symmetry discussed here and that for Calabi-Yau spaces we can
call the symmetry discussed here ``exceptional mirror symmetry''.

Note that the term ``mirror symmetry'' originates from the relationship
between
Hodge diamonds in mirror Calabi-Yau spaces. 
Even though we choose the name mirror symmetry
for the analogous symmetry discussed here for manifolds of exceptional
holonomy, the analogue of the Hodge diamonds in the ``mirror'' target
spaces are not related in any similar fashion \cite{ba1}.
\newpage
\section{$T$-duality as Mirror Symmetry for Joyce $7$-Orbifolds.}

In this section we will study Type IIA (or IIB) string theory defined 
on a class
of
Joyce orbifolds. Some of the orbifolds in this class (perhaps all) 
can be desingularised
to give smooth manifolds of $G_2$ holonomy.
However, regardless of whether some
``classical'' intepretation for these target spaces exists, these 
orbifolds are well
defined in string theory. This class of orbifolds will be defined as
orbifolds of the form 
${\bf T^7}/{\bf \Gamma}$, where ${\bf \Gamma} \cong {{\bf Z_2}^n}$.
Three of these ${\bf Z_2}$ 
generators act non-freely
on the torus; the remaining act freely and do not break any supersymmetry.

The simplest orbifold in this category may be defined as follows.

Consider the seven-torus, ${\bf T^7}$, with coordinates $x_1, x_2,....x_7$. Define
the ${\bf Z_2}^3$ orbifold group ${\bf \Gamma}$, with generators $\al,\bt,\ga$
as follows:
 
\be 
\alpha(x_i) = (-x_1,-x_2,-x_3,-x_4,x_5,x_6,x_7)
\ee
\be
\beta(x_i) = (-x_1,-x_2,x_3,x_4,-x_5,-x_6,x_7)
\ee
\be
\gamma(x_i) = (-x_1,x_2,-x_3,x_4,-x_5,x_6,-x_7)
\ee

In \cite{ba1}, we showed that under the $T$-duality transformation which
inverts the radii of the ${\bf T^3}$ with coordinates ${x_2},{x_3},x_5$,
Type IIA (IIB) theory on this orbifold is mapped to Type IIB (IIA) theory
on the same orbifold\ft{by the same orbifold we mean the orbifold
with the same generators, but of course with $T$-dual
radii.}, 
but with a certain ``amount'' of discrete torsion
turned on. Equivalence between these dual theories was checked to genus
$g$ in the string path integral. 
Let us denote the mirror transformation as $T_{235}$.
This is the first
of two types of mirror transformations we will discuss in $G_2$ holonomy
compactifications of Type II theories.

The assertion that the two theories 
are equivalent follows from the following formula
for how the genus $g$ path integral measure transforms under $T_{235}$:
\be
{\mu}_{g,\al} \longrightarrow (-1)^{{\sigma}_{\al}}.\epsilon.{\mu}_{g,\al}
\ee

Here, ${\mu}_{g,\al}$ is the path integral measure for the original
theory, at genus $g$. The genus $g$ Riemann surface has spin structure $\al$.
${\sigma}_{\al}$ is the parity of this Riemann surface. 
The factor corresponding to the
parity which appears in the ``mirror'' measure is responsible
for converting IIA to IIB and vice-versa. Finally, $\epsilon$ is the
discrete torsion in the ``mirror'' theory. This formula
was derived in \cite{ba1} using the techniques developed in \cite{vw} and
the explanation of why the proof of this formula is a proof of equivalence
between the two theories is also explained in \cite{vw}.
Following the notations of the
last section, let us denote the target space of the undualised theory as
${\bf X}$ and that of the mirror theory as ${\bf Y}$. Since the above formula applies
to all Joyce orbifolds of the form ${\bf T^7}/{{\bf Z_2}^n}$,
we are free to restrict
our attention to the subclass which
contains those which admit desingularisation
to manifolds of $G_2$ holonomy. We may do this in order
to discuss the implications of this ``mirror'' symmetry for the structure of
bona fide manifolds of $G_2$ holonomy. However our considerations also
apply to any of the more stringy target spaces which might exist in the class
of orbifolds we are considering.

If we take the original theory to be Type IIA on ${\bf X}$, then the moduli space
of the zero-brane\ft{All the branes we discuss in this paper
are D-branes. We drop the ``D'' for simplicity.} is ${\bf X}$ (here and
throughout this paper we make the assumption that all $T$-duality symmetries
are exact.).
Under $T_{235}$, this brane is mapped to the
$3$-brane of Type IIB theory wrapped around a supersymmetric ${\bf T^3}$
in ${\bf Y}$.
Because the target spaces ${\bf X}$ and ${\bf Y}$
differ only by the inclusion of
discrete torsion, their untwisted sectors are the same. Geometrically
this sector is associated with the elements of homology/cohomology of
${\bf T^7}$, which are invariant under the orbifold group. Because the mirror
transformation can be applied at the orbifold limit of the theory,
the superymmetric ${\bf T^3}$ around which the $3$-brane is wrapped should
correspond to a ${\bf T^3}$ submanifold of ${\bf T^7}$
invariant under the orbifold
isometry group. Moreover, with the choice of coordinates given above,
this ${\bf T^3}$
has local coordinates ${x_2},{x_3},x_5$. Let us denote this torus
by ${\bf T^3}_{235}$. In fact, at a given point in moduli space,
${\bf T^3}_{235}$ is one
of only seven supersymmetric $3$-tori present in the untwisted sector of the
orbifold. An easy way to see this is that there exists on 
${\bf T^7}/{{\bf Z_2}^n}$ a covariantly constant $3$-form, $\phi$.
Its components are given by\ft{$dx_{ijk}$ means 
${dx_{i}{\wedge}dx_{j}{\wedge}dx_{k}}$.}:
\be
{\phi} = dx_{127} + dx_{136} + dx_{145} + dx_{235} - dx_{246} + dx_{347}
+ dx_{567}
\ee

$\phi$, being $G_2$ invariant encodes the holonomy structure of the target
space. The
restriction of ${\phi}$ to any of the invariant $3$-tori in ${\bf T^7}$ 
is the volume form on that $3$-cycle.
Such cycles are minimal volume \cite{cal} and hence supersymmetric (BPS).

One can consider $T$-dualising these other tori, and one indeed
finds other examples of mirror pairs of theories. It would be interesting
to compute the Betti numbers of these other mirror pairs of target spaces.

Since in these examples one has a concrete understanding that the two
theories are mirror, it must be true that the moduli space of the $3$-brane
wrapped around ${\bf T^3}_{235}$ in ${\bf Y}$ is ${\bf X}$.
This result gives a prediction for the dimension of the space
of supersymmetric deformations of 
$3$-tori in manifolds of $G_2$ holonomy, which
we will discuss in sections five and six. 

We now give an example which
involves $T$-dualising a supersymmetric four-torus in the same class
of Joyce orbifolds. This is the second type of
mirror symmetry transformation which we mentioned earlier.
Consider the supersymmetric four-torus, 
${\bf T^4}_{1467}$,
where the subscript indicates the coordinates of this cycle.

Following the techniques in \cite{vw},
it can be shown that under the $T$-duality transformation ($T_{1467}$)
which inverts the radii of this torus, the
path integral measure of the Type IIA or IIB theory transforms as follows:

\be
{\mu}_{g,\al} \longrightarrow \epsilon.{\mu}_{g,\al}
\ee

This formula shows that for all the target spaces in the category we
are considering, Type IIA (or IIB) theory on the orbifold without
discrete torsion transforms into Type IIA (or IIB) theory on the same
orbifold, but with discrete torsion turned on. Furthermore, if we began
with Type IIA theory on ${\bf X}$, the zero-brane is mapped to a $4$-brane in IIA
theory on ${\bf Y}$
wrapped around ${\bf T^4}_{1467}$.
The moduli space of this brane is ${\bf X}$. In sections five and
six we discuss this moduli space. 
Since the above results apply to a large number of
mirror target spaces, this is strong evidence that the SYZ argument
can be successfully applied
to mirror manifolds of $G_2$ holonomy. 
As was the case for the supersymmetric $3$-tori,
in this class of orbifolds there also exist seven supersymmetric $4$-tori
in the ``untwisted sector'' of the orbifold, of which ${\bf T^4}_{1467}$ is
just one example. In fact these seven $4$-tori are just 
the Hodge-Poincare duals of the seven supersymmetric $3$-tori.
$T$-dualising these other six $4$-tori gives further examples
of mirror target spaces.

It is clear from our introductory discussion that these mirror 
transformations
are direct analogues of the mirror symmetry transformations in
${\bf T^m}/{{\bf Z_2}^n}$ Calabi-Yau orbifolds.

\section{Joyce $8$-orbifolds.}

We can also give examples of mirror target spaces of $Spin(7)$ holonomy.
These are based on the Joyce orbifolds considered in \cite{J3}.
The class of such orbifolds that we will consider are all orbifolds of the
form ${\bf T^8}/{{\bf \Gamma}}$. Here ${\bf \Gamma} \cong {{\bf Z_2}^n}$.
Four of the ${\bf Z_2}$
generators of ${\bf \Gamma}$ each break half the supersymmetries present in
compactification on ${\bf T^8}$, ie together
they preserve $1/16$ of the original
supersymmetry. The remaining generators act freely on the torus and
break no supersymmetry. The simplest example in this category can be defined
as follows. Compactify the example above (eqs. $(1)-(3)$) on a further
${\bf S^1}$ with coordinate $x_8$, 
and orbifold the theory by a further ${\bf Z_2}$ isometry which acts in the
following way:
\be
\delta(x_i) = (x_1,x_2,x_3,x_4,-x_5,-x_6,-x_7,-x_8)
\ee

We are thus considering the orbifold of ${\bf T^8}$ with coordinates
$(x_1,...x_8)$ and generators $(\al,\bt,\ga,\delta)$. In the untwisted
sector of any of the orbifolds in this class, there exist $14$
supersymmetric $4$-tori. However, Hodge duality interchanges
the volume forms of these tori, giving seven pairs of dual
$4$-tori. The volume forms are given by the components
of the $Spin(7)$ invariant $4$-form. Our choice
of coordinates coincides with that of Joyce \cite{J3}, and the components
of this $4$-form are given in \cite{J3}. 
Let us denote these $14$ supersymmetric $4$-tori by ${\bf T^4}_i$. 

It may then be checked that upon $T$-dualising ${\bf T^4}_i$, 
the path integral
measure of the Type IIA or IIB theory transforms in the following way:
\be
{\mu}_{g,\al} \longrightarrow {\epsilon}{_i}.{\mu}_{g,\al}
\ee

Here, ${\epsilon}_i$ is the discrete torsion associated with the $i$'th
mirror transformation. The discrete torsion in each of the fourteen cases
can be calculated following the methods of \cite{vw}. This formula shows
that under the fourteen mirror symmetry transformations Type IIA (or IIB)
theory on ${\bf X}$ transforms into Type IIA (or IIB) theory on ${\bf Y}$;
where ${\bf X}$ and
${\bf Y}$ are both orbifold
target spaces from the class of $8$-orbifolds we are
restricting our attention to. As orbifold conformal field theories,
${\bf X}$ and
${\bf Y}$ differ by the discrete torsion factor ${\epsilon}_i$. Clearly, by 
considering the zero-brane whose moduli space is
${\bf X}$ (${\bf Y}$), under all these
mirror transformations, this brane is mapped to a $4$-brane wrapped around
${\bf T^4}_i$. This result thus makes a prediction about the moduli space
of supersymmetric deformations of a supersymmetric $4$-torus in a manifold
of $Spin(7)$ holonomy, which we will consider shortly.

\section{Toroidal Fibrations and Mirror Symmetry.}

In this section we study the
generalisation of
the Strominger, Yau, Zaslow argument to mirror
manifolds of exceptional holonomy. 
Since the Type II theories have $N$$=$$2$ supersymmetry in ten dimensions,
when compactified on a manifold of special holonomy, the resulting $(10-d)$-
dimensional theory also has $N$$=$$2$ supersymmetry. In particular, this means
that Dirichlet $p$-branes which are reduced from ten dimensions break half
of these supersymmetries and are BPS states. Because the perturbation theories
of the two mirror theories are directly related (as is the case for mirror
Calabi-Yau threefold compactifications), we can compare the BPS soliton
spectra in both theories. In particular we can expect
the moduli spaces of BPS branes
exchanged by the mirror transformation to be identical.
(In order for this to be true the mirror symmetry must be
an exact non-perturbative symmetry. The $T$-dualities of
the previous sections are exact perturbatively. However we
expect $T$-dualities to be exact dualities of the full
theory and not just an artefact
of string perturbation theory. 
This structure is of course required
by $U$-duality \cite{cp}). In general, as pointed
out in \cite{syz} there will be world-sheet instanton corrections to the
moduli spaces of wrapped branes.
However in this paper we will neglect these corrections as we
wish to focus on the general structure predicted by exceptional
mirror symmetry. As in
\cite{syz} these
corrections will be important in understanding the precise details
of the picture which emerges.

Let us now consider applying the SYZ argument with the assumption
that a pair of mirror Type II theories exists.
We can begin by considering Type IIA theory on 
${\bf {R^{10-d}}}{\times}{\bf X_d}$.
The starting point for applying the SYZ argument is the observation that
the moduli space of the zero-brane, dimensionally reduced from ten
dimensions, is ${\bf X_d}$. We then look for this zero-brane in the mirror
theory, which is either Type IIA or IIB theory on ${\bf Y_d}$. In the case
where the mirror theory is Type IIA on ${\bf Y_d}$, we can not identify
the dimensionally reduced zero-brane as the brane we are looking for
because its moduli space is ${\bf Y_d}$ and ${\bf Y_d}$ and ${\bf X_d}$ are different
manifolds.
We may therefore conclude that the zero-brane in the string theory
on ${\bf Y_d}$ is a 
Dirichlet $p$-brane which is wrapped around a supersymmetric
$p$-cycle \cite{be},
$C_p$ $\subset$ ${\bf Y_d}$. Because the two theories are mirror, the
moduli space of this wrapped $p$-brane is ${\bf X_d}$.

Let us consider the structure of the moduli space of the wrapped $p$-brane
world-volume theory. We are lead to consider the ``compactification''
of this $p+1$ dimensional theory on $C_p$, to $0+1$ dimensions.
The theory contains a $U(1)$ gauge field.
Reducing this gauge field on $1$-cycles
of $C_p$ gives ${b_1}(C_p)$ scalars in the reduced world
volume theory. 
With all other moduli fixed, the moduli space of these scalars
is the torus ${\bf T^{b_1}}$. The other moduli in the theory correspond to
supersymmetric, normal deformations of $C_p$ within ${\bf Y_d}$. 
We will call this
moduli space the space of supersymmetric deformations and denote it by 
${\bf B_c}$,
where $c$ is its dimension. At a generic\ft{We say generic because there
will in general exist points on the base at which the fiber 
degenerates.} point on ${\bf B_c}$, we have
a torus ${\bf T^{b_1}}$. We can therefore see that the moduli space
of the wrapped $p$-brane theory has the structure of a fibration with
base ${\bf B_c}$ and fiber ${\bf T^{b_1}}$. This is by no means a precise statement.
However, intuitively this is the natural picture which emerges.
This structure has been useful in similar,
related discussions of wrapped branes \cite{syz,dtop}.
Furthermore, the later arguments that we will give
provide further
evidence that this is indeed
the structure of the moduli space at generic points. For now we will 
assume that this fibration picture of the moduli space is correct. 

Since the space in question is ${\bf X_d}$,
we require 
\be
c + {b_1}(C_p) = d
\ee

We can now restrict our attentions to the cases when both ${\bf X}$ and
${\bf Y}$ have exceptional holonomy.

Since $C_p$ is a supersymmetric cycle, it is a calibrated submanifold
of ${\bf Y_d}$. 
This means that there exists on ${\bf Y_d}$ a closed exterior $p$-form
(${\phi}_p$) whose restriction to $C_p$
is the volume form of $C_p$ \cite{cal}.
On $7$-manifolds of $G_2$ holonomy there exists a covariantly constant 
$3$-form
whose Hodge dual is a covariantly constant $4$-form.
The two natural classes of
supersymmetric cycles in $G_2$ holonomy manifolds are therefore 3-cycles
and 4-cycles. In fact, these are the only known classes of supersymmetric
cycles on such manifolds.
For compactification on manifolds of $Spin(7)$ holonomy, the only known
class of supersymmetric cycles are $4$-cycles, which are calibrated by
the covariantly constant $4$-form, which encodes the holonomy structure of
the manifold. The fact that the calibrated submanifolds of
manifolds of exceptional holonomy are supersymmetric cycles was shown in
\cite{dtop,mor}.

Let us consider the $4$-cycles
in the case when both ${\bf X}$ and ${\bf Y}$ are $7$-manifolds of $G_2$ holonomy.
Mclean \cite{mclean} has shown in this case
that
\be
c = {b_2^+}(C_4)
\ee
where ${b_2^+}$ denotes the number of self-dual harmonic $2$-forms.

We can now consider equation $(9)$. This becomes

\be
{b_2^+}(C_4) + {b_1}(C_4) = 7
\ee

The only $4$-manifold we know, which satisfies
this constraint, is the $4$-torus ${\bf T^4}$. 
It is not known to us whether
or not this is the only $4$-manifold for which this is true. However,
assuming that this is the case, our zero-brane in Type IIA theory on
${\bf X_7}$ may be identified with a $4$-brane wrapped around a supersymmetric
${\bf T^4}$ in ${\bf Y_7}$.
Furthermore, the moduli space of this $4$-brane (which
we identify with ${\bf X_7}$) is a ${\bf T^4}^{\prime}$
fibration over ${\bf B_3}$ (where ${\bf T^4}^{\prime}$ is the $4$-torus
which is the moduli space of the gauge connections). Moreover
the zero-brane in the Type IIA theory on ${\bf Y_7}$ can be identified
with a $4$-brane in the theory on ${\bf X_7}$, wrapped around another 
supersymmetric
${\bf T^4}$.
We therefore learn that both ${\bf X_7}$ and ${\bf Y_7}$ contain
supersymmetric $4$-tori and are (at least locally)
fibred by these $4$-tori.

We can then
identify the mirror symmetry transformation as the $R \rightarrow 1/R$
$T$-duality transformation on the ${\bf T^4}$ fibres, since we
know that the zero-brane and $4$-brane are interchanged under such a
transformation. This implies that the ${\bf T^4}$ fibers of ${\bf X}$ 
are $T$-dual
to those of ${\bf Y}$. The results of section three provide 
strong evidence that the SYZ argument has been successfully applied here.

To summarise the results so far,
we have learned that in Type IIA compactification
on a $7$-manifold of $G_2$ holonomy (${\bf X}$), the existence of a mirror
Type IIA theory on the mirror manifold of $G_2$ holonomy (${\bf Y}$) leads to
the conclusion that both ${\bf X}$ and ${\bf Y}$ admit supersymmetric 
${\bf T^4}$
fibrations. The mirror transformation is $T$-duality on the ${\bf T^4}$
fibers. Note that strictly speaking
this conclusion may only be drawn if ${\bf T^4}$ is the
only $4$-manifold which satisfies equation $(11)$.

So what about the $3$-cycles in the $G_2$ case and the $4$-cycles in the
$Spin(7)$ case ? Unfortunately, in these two cases, $c$ is difficult
to determine, because it is not known to be a topological invariant of the
supersymmetric cycle. In fact, in these cases, $c$ is the number of
harmonic spinors with values in a particular vector bundle, whose
definition may be found in \cite{mclean}. In order to deal
with this situation we will make further assumptions.

Let us first discuss the general picture which is emerging. For the case
of Calabi-Yau $n$-folds and the case of $7$-manifolds of $G_2$ holonomy
which we have just discussed,
the SYZ argument suggests that when such manifolds have mirrors, they are
fibered by supersymmetric tori. The ultimate reason that this is so is that
mirror symmetry is $T$-duality. 
Thus, the picture
emerging is that of mirror manifolds with supersymmetric cycles which fiber
the whole manifold.
It is natural to hope that a similar
story will emerge for the $3$-cycles on $7$-manifolds of $G_2$ holonomy
and for the $4$-cycles on manifolds of $Spin(7)$ holonomy.

With this in mind, consider the case in which the supersymmetric cycle $C_p$,
locally fibers the whole manifold ${\bf Y_d}$. In this case, the
space of supersymmetric deformations has dimension
\be
c = d - p
\ee

Let us assume that this is the case and consider equation $(9)$
again. This now says that
\be
b_1(C_p) = p
\ee

This equation has a natural solution: $C_p$ is a flat $p$-torus ${\bf T^p}$.
What this shows is that
if there exists a supersymmetric cycle $C_p$ in ${\bf Y_d}$ 
which locally fibers
all of ${\bf Y_d}$, 
then the moduli space of the $p$-brane wrapped around this
cycle has the same dimension as ${\bf X_d}$ if $C_p$ $\cong {\bf T^p}$. 
Thus if the
original zero-brane is to be identified with a $p$-brane wrapped around a
cycle which locally fibers all of ${\bf Y_d}$, then this identification
is possible if this cycle is a $p$-torus.

This simple result is quite remarkable since this is the general
feature which appears to emerge from the SYZ argument which ultimately
results in the statement that mirror symmetry is $T$-duality on such
tori.

It follows that if the dimension of ${\bf B_c}$ for a supersymmetric
${\bf T^3}$ (resp. ${\bf T^4}$) 
submanifold of a manifold of $G_2$ holonomy (resp.
$Spin(7)$ holonomy) is indeed four (resp. four), then by the SYZ argument
we should conclude that manifolds of $G_2$ ($Spin(7)$) holonomy which
have mirrors admit ${\bf T^3}$ (${\bf T^4}$) fibrations.

The SYZ argument given in this section has
been applied completely independently to the results of the previous
sections; we simply assumed the existence of mirror compactifications
of the Type II theories, according to the definition given earlier.
If we now consider our earlier results then we see that they provide
strong evidence for the conclusions drawn here. In particular, in the case
of supersymmetric 3-tori (4-tori) in manifolds of $G_2$ ($Spin(7)$) holonomy,
the results of sections three and four require the space of supersymmetric
deformations, ${\bf B_c}$, to be four dimensional in each case. Combining
these results with those in \cite{syz,mor} leads to the general picture that
manifolds with $SU(n)$, $G_2$ or $Spin(7)$ holonomy for which
a mirror exists admit supersymmetric ${\bf T^p}$ fibrations with
$p$ the dimension of the fiber varying from case to case. Moreover,
mirror symmetry is $T$-duality on these toroidal fibers.

\newpage

\section{Moduli Space Structure.}

The results of the previous sections have suggested a particular
fibration structure for manifolds of exceptional holonomy. Furthermore
the moduli spaces of certain wrapped D-branes should also share these
properties. In this section we will again appeal to the simplifying
features of orbifolds. The target spaces we have considered here
contain manifolds of exceptional holonomy which possess an orbifold
limit. In particular we can expect the orbifold, viewed as a space
in classical geometry to inherit this fibration structure. This is
because the arguments we have given apply in principle at any point
in the full moduli space of the string compactification, and the
``blowing up'' modes are moduli in this space. By considering
the classical orbifolds themselves we will indeed
see that the correct structure
emerges.

We begin with one of the Type II theories on a $d$-torus,
${\bf T^d}$ $\cong$ ${{\bf T^p}}{\times}{{\bf T^{d-p}}}$. 
Now we take an orbifold
of this theory.
Denote the finite orbifold isometry
group as ${\bf \Gamma}$. 
In general, each element of ${\bf \Gamma}$ will have a piece
which acts on ${\bf T^p}$ and a piece which acts on 
${\bf T^{d-p}}$. We
may write ${\bf \Gamma} \cong {h_p}.{h_{d-p}}$, where the subscripts denote
the pieces of $\Gamma$ which act on the tori of the corresponding dimension.

We will denote the orbifold we are considering as 
\be
{\bf M_{\Gamma}} = 
({\bf T^p}{\times}{\bf T^{d-p}})/({h_p}.{h_{d-p}}).
\ee
We will assume that the volume
form of ${\bf T^p}$ is preserved by ${\bf \Gamma}$ so that 
${\bf T^p}$ is a supersymmetric
cycle in the orbifold. Since the Hodge dual of this form is the volume
form of ${\bf T^{d-p}}$, ${\bf T^{d-p}}$ 
is also preserved by the orbifold group. Away
from the fixed points of ${h_{d-p}}$ in ${\bf T^{d-p}}$, 
we can view the whole
space as a fiber bundle with fiber 
${\bf T^p}$ and base ${\bf T^{d-p}}/{h_{d-p}}$ 
\cite{sen}. This description breaks down at the the fixed points in 
${\bf T^{d-p}}$
where the fibre ``degenerates'' to ${\bf T^p}/{h_p}$. 

The description we have
just given applies to the Joyce $7$- and $8$-orbifolds of the preceding
section. Moreover, in the case of the $7$-orbifolds the {\it only} 
${\bf \Gamma}$
invariant cycles are $3$-tori and $4$-tori (apart
from the zero cycle and the fundamental cycle). 
Similarly, the only ${\bf \Gamma}$ 
invariant $4$-cycles in the $8$-orbifolds are $4$-tori. Thus, away from the
``degeneration points'', the $7$-orbifolds have the structure of fibrations
with supersymmetric ${\bf T^3}$ (or ${\bf T^4}$) fibres and 
${\bf T^4}/{h_4}$ (${\bf T^3}/{h_3}$) base.
Similarly, the $8$-orbifolds have the structure of fibrations with
supersymmetric ${\bf T^4}$ fibres. Thus the classical orbifolds themselves
have the structure predicted by our application of the SYZ argument. We
expect that desingularisation of the orbifold leaves this structure intact
although it is difficult to make a precise statement about this.
We will now apply an argument which shows that the moduli space
of the world volume theory of a $p$-brane wrapped around these fibers has
precisely the same structure.

Consider the adiabatic limit in which 
${\bf T^{d-p}}/{h_{d-p}}$ is very large,
or equivalently that ${\bf T^p}$ 
varies slowly over the base. In this adiabatic
limit we can consider a $p$-brane wrapped around the supersymmetric
fiber ${\bf T^p}$. In this limit, the brane is generically away from the
``degeneration'' points. Thus, as far as the adiabatic brane is concerned,
it is wrapped around ${\bf T^p}$ and moves on ${\bf R^{d-p}}$. 
In this limit, under
$T$-duality on ${\bf T^p}$, 
the $p$-brane is mapped to a zero brane in Type IIA
theory compactified on the $T$-dual torus to ${\bf T^p}$. 
As time goes on the wrapped brane eventually discovers 
the space of all of the normal, supersymmetric deformations. 
Away from the degeneration points where the cycle degenerates,
the moduli space of the $p$-brane theory is thus a
fiber bundle with base ${\bf T^{d-p}}/{h_{d-p}}$.
The fibre in the moduli
space is a $p$-torus ${\bf T^p}^{\pr}$. 
This moduli space is $d$-dimensional if we
assume that the dimension does not change at the degeneration points. 
This is
a natural assumption if we require the physics to remain non-singular.

Clearly, by applying this adiabatic argument to a brane wrapped around a
supersymmetric cycle in any of the Joyce orbifolds of the preceding section,
we see that its moduli space has the same structure
as the orbifolds themselves. 
This gives us further confidence in our 
application of the SYZ argument since we indeed wanted to identify the
moduli space of the brane itself as a Joyce orbifold. 
Now let us consider the cases in which the target space orbifold 
admits a desingularisation
to a manifold of exceptional holonomy.
Clearly, desingularising
the target space orbifold induces a desingularisation of the 
wrapped brane moduli
space orbifold via the $T$-duality/mirror transformation. We thus find
that both the target space and the wrapped brane moduli space are manifolds
of exceptional holonomy.
In fact by considering Type IIA theory on ${\bf M_{\Gamma}}$ and applying the
relevant $T$-duality/mirror transformation, the moduli space of the relevant
wrapped D-brane is topologically the same classical orbifold as
${\bf M_{\Gamma}}$. This statement may be verified by noting that the
action of ${\bf \Gamma}$ on the torus in the $p$-brane moduli space
is precisely the same as its action on the torus in ${\bf M_{\Gamma}}$.

\section{Summary and Outlook.}

We have applied the SYZ argument to mirror Type II theories on
manifolds of exceptional holonomy. This has lead us to the following
conclusions: If Type IIA theory on a manifold of $G_2$ holonomy is
mirror to Type IIA (IIB) theory on a mirror manifold of $G_2$ holonomy, then
the mirror manifolds admit supersymmetric ${\bf T^4}$ 
(${\bf T^3}$) fibrations;
if Type IIA (or IIB) theory on a manifold of $Spin(7)$ holonomy has
a mirror, then the mirror manifolds admit supersymmetric 
${\bf T^4}$ fibrations.

Following our previous work \cite{ba1}, we have given examples of all three
of the above types of mirror transformations which apply to fairly large
classes of target spaces of exceptional holonomy. These transformations
are further evidence for the above conclusions. Furthermore, in the last 
section we showed that all the target spaces in this class of examples
have the above fibration structure; moreover, the relevant wrapped $p$-brane
moduli space does so also.

Combining these results with those of \cite{syz,mor} we can see that there
is a general picture in which manifolds which admit mirrors admit
supersymmetric fibration by $n$-tori, where $n$ varies from example to
example.

We believe that mirror symmetry for manifolds of exceptional holonomy
certainly deserves further study; both in string theory and mathematically.
In the latter area, perhaps some more rigorously understood examples
can be constructed along the lines of the construction in \cite{wilson}.
In the case of string theory, mirror symmetry for Calabi-Yau
threefolds is defined perturbatively at the level of conformal field
theory. In the case of Calabi-Yau
target spaces the mirror symmetry is directly related to the representations
of the superconformal algebra that underlies the string theory propagation on
such spaces. 

In \cite{sv} a generalised mirror conjecture was made. 
Applied to superstrings on manifolds of exceptional holonomy,
this conjecture is related to the superconformal
algebras associated with string propagation on target spaces
of exceptional holonomy. As noted in \cite{ba1} the mirror
symmetries discussed here provide examples for which the generalised
mirror conjecture holds true.
It may be that a more
precise definition of exceptional mirror symmetry could be
given at the level of the superconformal field theories written down in
\cite{sv}, although it is not at all obvious to us if this is possible.

Finally we would like to mention that evidence for ${\bf T^3}$ fibered manifolds
of $G_2$ holonomy has recently appeared in \cite{ru} in the context of
constructing the $U$-manifolds of \cite{U},
although the precise details of this
construction have not been given.

\section{Acknowledgements.} We would like to thank M. O'Loughlin for
comments on an earlier version of this paper and J.M. Figueroa-O'Farrill
for discussions. We would also like to thank PPARC, by whom this work is
supported.

\end{document}